\documentstyle[preprint,prl,aps]{revtex}
\begin{document}
\tighten
\title{Pearling and Pinching: Propagation of Rayleigh Instabilities}
\author{Thomas R. Powers$^{1,2}$\cite{newaddress} and
Raymond~E. Goldstein$^{1}$\cite{newaddress}}
\address{$^{1}$Department of Physics, Joseph Henry Laboratories,
Princeton University, Princeton, NJ 08544}
\address{$^{2}$NEC Research Institute, 4 Independence Way, Princeton, NJ
08540}

\date{\today}
\maketitle
\begin{abstract}
A new category of front propagation problems is proposed
in which a spreading instability evolves through a singular configuration
before saturating.  We examine the nature of this
front for the viscous Rayleigh instability of a column of one fluid immersed
in another, using the marginal stability criterion to estimate the
front velocity, front width, and the selected wavelength
in terms of the surface tension and viscosity contrast.
Experiments are suggested on systems that may display this
phenomenon, including droplets elongated in extensional flows,
capillary bridges, liquid crystal tethers,
and viscoelastic fluids.  The related problem of propagation in
Rayleigh-like systems that do not fission is also considered.

\end{abstract}

\pacs{PACS numbers: 47.20.-k, 68.10.-m, 02.30.Jr, 47.54.+r}

Front propagation problems are often divided into
two broad classes:  the invasion of one stable state by another,
and the invasion of an unstable state by a stable one \cite{CrHo}.
Here we point out
the possibility of a third, qualitatively different type of front
propagation:  the invasion
of an unstable state by one that develops discontinuously
into its final configuration.  Although we expect this class
to be very broad,
we focus on one particular example, the Rayleigh instability, wherein
a cylindrical body of fluid longer than its
circumference is unstable to breakup into droplets \cite{Plateau}.
This is a {\it linear} instability \cite{Strutt} triggered by infinitesimal
perturbations of sufficiently long wavelength.

Recent work has shown Rayleigh instabilities in many
experimental systems: laser-tweezed cylindrical
lipid vesicles \cite{BarZiv},
liquid crystal tethers embedded in a polymer matrix \cite{Mather},
shrunken polymer gels \cite{MT}, and
columns of viscoelastic fluid \cite{VE}.
Traditional experiments display an instability
developing uniformly along the cylinder \cite{expts}.
However, experiments on the ``pearling instability''
of laser-tweezed membranes
reveal a different scenario---{\it front propagation}.
The peristaltic shape deformation spreads out from the laser spot with
a constant velocity\cite{slowing_down}.
The emerging explanation \cite{PNS,Granek,GNPS} for this
instability is that it is driven by the
tension induced in the membrane by the laser trap.
Fissioning is ultimately prevented by the membrane elasticity.

Motivated by the phenomenology of the pearling instability,
we ask,
``Can the Rayleigh instability
develop as a propagating front?''
(as in Fig. \ref{fig1}).
Propagation in this context is problematic.
As a droplet pinches off, the two tips of the
broken neck recede from the pinching point;
if the retracting neck overtakes the front, propagation will be spoiled.
Classic experiments on
the breakup and relaxation of elongated droplets
\cite{SBL} reveal that this competition
depends on the viscosity contrast of the two fluids.
This retraction and topological rearrangement of interface
is absent in
overdamped systems known to have propagation, such as
dendritic growth, viscous fingering, and
reaction-diffusion systems.
The mathematical structure of our problem differs from
those just mentioned as well, with a global
conservation law from fluid incompressibility, nonlinearities from
the two principal surface
curvatures, and singular behavior near pinching.
The former two features
are also common to Rayleigh-like systems that, like membrane tethers,
do not fission.

The novel aspects outlined above are the very features
that complicate both analytical and numerical analysis of our
problem.
As a first step, we apply the intuition gained from
front propagation problems in local partial differential equations
(PDEs) to the Rayleigh problem, with the hope that these estimates
will serve as a guide for further investigation.

With the exception of the one-dimensional nonlinear
diffusion equation \cite{AW},
there are no rigorous analytic results for front propagation in PDEs.
On the other hand, in many cases the marginal stability criterion (MSC)
\cite{DeeL} correctly predicts the front properties.
We apply the MSC to the Rayleigh instability
to make concrete predictions for front velocity,
front sharpness, and selected wavelength, and
suggest some experimental systems where this phenomenon may be seen.

The three categories of front motion are illustrated by
generalized Fisher-Kolmogorov (FK) equations\cite{Fish},
\begin{equation}
\label{fisher}
u_t=-{\delta {\cal E}\over {\delta u}},
\end{equation}
where ${\cal E}=\int {\rm d}x (u_x^2/2-{\cal V}(u))$, with three types of
potentials shown in Fig. \ref{fig2}.
Recall the standard mechanical analogy for traveling-wave solutions:
the ansatz $u(x,t)=f(x-vt)$ reduces (\ref{fisher}) to Newton's equation
for a particle of coordinate $u$ at time $z=x-vt$ in a
potential ${\cal V}(u)$ with friction coefficient $v$.
In case (a) there is a unique $v$
that allows the front
to connect the two locally stable states.
When a stable state invades an unstable one (b),
the particle reaches
$u=0$ at $z=\infty$ for every friction coefficient $v>0$.
However, for sufficiently localized initial conditions, there is a unique
front velocity that appears asymptotically, and is correctly predicted by
the MSC.
Finally, (c) displays a situation loosely analogous to the
new category.  A potential of this form, {\it e.g.} ${\cal V}(u)=
u^2/2+u^4/4$, leads to a finite-time blow-up of $u$ \cite{blowup}.
Although this model does not describe a topological transition,
it is like the Rayleigh problem in which the growing instability
has a finite-time singularity.

This mechanical analogy suggests that case (c) would, like (b), possess a
family of possible velocities.  With this motivation, we examine the
Rayleigh problem using the {\it linear} MSC, which yields
predictions directly from the linear growth rate $\omega(q)$ \cite{nlmsc}.
The front velocity $v^*$ and the shape of the front's
leading edge,
$u(x,t)\sim\exp(\omega(q^*)t+i q^* x)$, are given
by the relations \cite{DeeL}
\begin{equation}
v^*={{\rm Re} \omega^*\over {\rm Re} q^*},\ \
{\rm Im} {\partial\omega^*\over\partial q}=0,
\ \ v^*={\rm Re}{\partial\omega^*\over\partial q},
\end{equation}
where $\omega^*=\omega(q^*)$ and $q^*=q'+iq''$ is complex.
In the standard MSC treatment, the selected wavenumber $q_0$ of the
saturated pattern is
different from the selected wavenumber $q'$ in the
leading edge of the front; $q_0$ is deduced by working in the rest frame of
the front and assuming nodes are conserved
as they pass through the front and into the saturated pattern.
We do not expect $q_0$ to be relevant to the Rayleigh problem since the
evolution of the interface is not continuous.

The conservation of fluid in the Rayleigh problem is like
that appearing in the invasion of the disordered phase of a
block copolymer by the lamellar phase\cite{Goldenfeld}, with global
conservation of the mean composition. In contrast to the FK equation, this
leads to a dynamics of the form $u_t+j_x=0$, with the
flux $j=-{\partial_x}(\delta {\cal E}/{\delta u})$, or
\begin{equation}
u_t={\partial^2\over\partial x^2}{\delta {\cal E}\over {\delta u}}~.
\label{blcopol}
\end{equation}

Consider now a thread of fluid of radius $R$ and
viscosity $\eta^-$ in another fluid of viscosity $\eta^+=\eta^-/\alpha$,
with interfacial tension $\Sigma$.
The growth rate of
a volume-preserving axisymmetric
sinusoidal perturbation with wavenumber $q=k/R$ follows
from the Stokes equations
$\eta^{\pm}\nabla^2 {\bf v}^{\pm} = \bbox{\nabla} p^{\pm}$,
subject to the boundary conditions of no-slip, no tangential stress, and
the Laplace law for the jump in normal stress,
$(\sigma^+_{ij}-\sigma^-_{ij})n_j=-2\Sigma H n_i$.
Here, $n_i$ is the outward surface normal,
$H$ is the mean curvature
$2 H=
{r_{xx}
{(1+r_x^2)^{-3/2}}}-{{r^{-1}(1+r_x^2)^{-1/2}}}~$,
and the stress tensors are
$\sigma^\pm_{ij}=
\eta^\pm(\nabla_i v^\pm_j + \nabla_j v^\pm_i) -p^\pm \delta_{ij}$.
The growth rate is \cite{Tomo}
$\omega(k,\alpha)={(\Sigma/{R\eta^+})} \Lambda(k,\alpha)(1-k^2)~$.
\label{tomo_growth}
The dynamical factor $\Lambda(k,\alpha)$ (too lengthy to quote here)
accounts for the dissipation, and the factor $\Sigma(1-k^2)$ is associated
with the energy of a distortion.

We gain intuition about $\Lambda(k,\alpha)$ from its
small $k$ form:
\[
\Lambda\sim\! {-k^2\left[-{1\over 8}+\alpha({1\over4}+{1\over 2}\gamma
-{1\over 2}\log 2)+{1\over 2}\alpha\log k\right]
\over{\alpha-
k^2(\alpha-1)\left[{1\over 8}+\alpha({3\over 4}+
{3\over 2}\gamma-{3\over 2}\log
2)+{3\over 2}\alpha\log k\right]}}~,
\]
where $\gamma$ is Euler's constant.
For finite $\alpha$ the asymptotic limit as
$k\rightarrow 0$ is $\Lambda(k,\alpha)
\sim-{1\over2}k^2\log k$.  The logarithm is due to large-scale
flows in the outer fluid, while the two powers of
$k$ are analogous to the two $x$-derivatives in (\ref{blcopol}).
They can be understood intuitively
by noting that one power comes from the flux form of the dynamics,
$\pi \partial_t r^2+\partial_x J=0$ \cite{GNPS}, with flux
$J\sim r^2 U$, where $U$ is the average axial velocity inside
the tube.  For small $k$ the Stokes equations imply
a Poiseuille flow of the inner fluid and thus the second power of $k$
through Darcy's law for the average velocity, $U\sim- r^2 p_x$.

If we consider
the case of $\alpha=\infty$ (no outer fluid), we recover the small
$k$ limit of Rayleigh's stability analysis\cite{Strutt}, which has a
non-zero growth rate at $k=0$.
The inner fluid has plug
flow since it does not have to entrain the outer fluid as
the shape of the boundary changes.  This in turn modifies the relationship
between the average velocity $U$ and the pressure so for small $k$,
$U_{xx}\sim p_x$\cite{papa}.

Fig. \ref{fig3} displays the results of the MSC applied to
Tomotika's growth rate
for a range of viscosity contrasts $\alpha=\eta^-/\eta^+$ \cite{second_branch}.
For example, the results for the castor-oil-eugenol
mixture immersed in silicone fluid ($\alpha=0.0315$)
used in \cite{Saville} are the following:
the fastest-growing mode
$k_{\rm max}=0.502$, $k'=q'R=0.404$,
$k''=q''R=0.291$, and $v^*=1.04 \Sigma/\eta^+$.
The MSC predicts a relatively sharp front;
if we measure the front width in units of $2\pi/k'$ we
see that it is generically a fraction of a growing bulge, from around
$0.25$ at $\alpha=1$ to about $0.1$ for large and small $\alpha$.
We note also that $v^*$ differs markedly from the naive estimate
$\omega(k_{\rm max},\alpha)/k_{\rm max}$. From these results we estimate the
distance
$(k'/{2\pi})(v^*/\omega(k'))$
the front moves in the time it takes a bulge to pinch off; it is
about twice the front width for the range of $\alpha$ shown, so
the droplets pinch off right behind the front.

Now we crudely estimate the retraction velocity of the bulge at the
end of a broken thread to see when we expect to find propagation.
Assuming a spherical
bulge, we balance the tension force with the drag
of a liquid sphere of radius $R$ \cite{Lamb} to deduce a velocity
$V=(\Sigma/\eta^+)(r/R)(\alpha + 1)/(2\alpha + 3),$
where  $r<R$ is the radius of the
neck attaching the bulge to the rest of the thread.
Note
that $V$ does not depend sensitively on the inner viscosity.
Comparing this expression with Fig. 3, we see that the front will outrun
retraction at small $\alpha$ but not for large $\alpha$.  Therefore, we
expect the MSC to describe front propagation at small $\alpha$ if it
applies at all.

These predictions are consistent with the experimental
results of \cite{SBL}, which concern the relaxation and breakup of an initially
elongated drop in an otherwise quiescent fluid.  It is found that
for large $\alpha$ the bulbs on the ends of the dog-bone shaped
droplet retract a substantial amount before they pinch off (if they pinch
off at all); for lower values of $\alpha$, the retraction is relatively
slower and the ends do not move as much before they pinch off.
As alluded to above, this behavior occurs
because the outer viscosity
limits the retraction velocity and the inner viscosity inhibits
pinching \cite{SBL}.  The experimental results at low $\alpha$ are
tantalizingly close to our conception of a propagating Rayleigh instability.
We note that the mechanism of ``end-pinching''
in which the end of the extended drop bulges and pinches off while
the rest of the drop is stationary \cite{SBL} is consistent with the MSC
prediction of a sharp front.  It remains to be seen if these disturbances
spread at a {\it constant} velocity.

Since the applicability of the MSC to our problem is speculative,
it is important to check our predictions.
A numerical computation of the interface evolution
in the fissioning Rayleigh problem is a challenging task.  As a first
step to test the internal consistency of the MSC we studied a
model PDE for a Rayleigh-like instability that approaches pinching,
with an energy functional
with an elastic contribution like that for membranes.
Using a Poiseuille flow approximation to relate the fluid flux
to the pressure gradient, we obtain
\begin{equation}
\partial_t r^2 = {1\over 4\eta}\partial_x\left(r^4\partial_x
{1\over 2\pi r}{\delta {\cal E}\over \delta r}\right)~,
\label{lubrication_eqn}
\end{equation}
with
${\cal E}=\int\! {\rm d}S\left\{\Sigma+(1/2)\kappa (2H)^2\right\}$
and $\kappa$ the bending modulus.
Note the natural progression from (\ref{fisher}) to (\ref{blcopol})
to (\ref{lubrication_eqn}).
Although (\ref{lubrication_eqn}) has a non-singular evolution,
it has the conservation laws and nonlinear structure
emphasized in the introduction.  This structure is similar
to thin-film dynamics \cite{Limat} in which
propagating Rayleigh-Taylor instabilities have been observed.
In \cite{GNPS} we found that (\ref{lubrication_eqn}) supported
front solutions; Fig.~\ref{fig4} shows close
agreement between the numerical solution of (\ref{lubrication_eqn})
and the MSC.  The discrepancy is likely due to the slow approach
of the velocity to its asymptotic limit \cite{DeeL} coupled with
computational limitations.
Note in Fig.~\ref{fig4} how as each bulge forms
it is pulled away from the thin tether, in accord with our argument
about tension and drag.

We now turn to possible experiments.
The examples we list here do not necessarily have a singular evolution;
they illustrate the other novel features of propagation
outlined in the introduction and raise as well the issue of the
mechanism by which pinching is inhibited in each case.
(i) In addition to droplets in extensional flow, another ideal
candidate for propagation with pinching is the
neutrally buoyant capillary bridge stabilized by an
electric field \cite{Saville}.
(ii) Mather {\it et al.}\cite{Mather} are studying a
shape instabilities of a thread of liquid crystal polymer
embedded in a polymer matrix, with a competition between tension and
nematic elasticity.  (iii) Matsuo and Tanaka \cite{MT} have stretched
a cylindrical gel between two glass plates and observed a
peristaltic instability in the presence of an appropriate solvent.
In addition to the long time scales, these systems have the advantage that
thermal fluctuations are small.  These facts may conspire
to allow one to see a propagating Rayleigh instability
in the cases without a
precisely controllable tension.
(iv) Jets of viscoelastic fluid, as in the work
of Renardy, {\it et al.} \cite{VE}, may also display
propagating Rayleigh instabilities.
(v) Finally, we expect there are examples of front propagation with
singular evolution that are not Rayleigh-like.
For example, in surface-tension-driven B\'enard
convection\cite{drain}, an initially flat layer of
liquid distorts to expose dry spots when it is heated from below.
We suggest that these dry spots can spread behind a front that
moves at constant velocity, rather analogous to those seen in the
Rayleigh-Taylor instability \cite{Limat}.

We have argued that front propagation in Rayleigh and Rayleigh-like
systems is qualitatively different from more familiar examples of
propagation.  The differences are both mathematical and physical:
the non-pinching Rayleigh-like systems have global conservation laws
and a particular nonlinear structure due to the two dimensional
nature of the interface, whereas the Rayleigh problem in addition
has singularities and the new
physical ingredient of retraction.
If these instabilities do indeed propagate, it
will be important to determine whether the propagation can
be described by the MSC calculations presented here
or if some new picture is needed.  There is a clear need for future
experimental and theoretical investigations.

We thank
R. Bar-Ziv, A.T. Dorsey, H. Levine, A.J. Liu,
P. Mather, E. Moses, P. Nelson, W. van Saarloos,
D. Saville, U. Seifert, H.A. Stone, and C. Wiggins for helpful
discussions.
Work at Princeton University was supported by NSF Presidential
Faculty Fellowship grant DMR 93-50227, and the A.P. Sloan Foundation (REG).

\begin{figure}
\caption[]{Propagation of the Rayleigh instability.}
\label{fig1}
\end{figure}

\begin{figure}
\caption[]{Potentials illustrating different classes of front propagation
in the mechanical analog for traveling-wave states.}
\label{fig2}
\end{figure}

\begin{figure}
\caption[]{Length scales and front velocity from the marginal stability
criterion.}
\label{fig3}
\end{figure}

\begin{figure}
\caption[]{Numerical results. Propagation of the Rayleigh instability
with an elastic cutoff (top). Front velocity as a function
of tension (bottom).  Solid line is MSC prediction.  Dashed line indicates
critical tension for onset of instability.}
\label{fig4}
\end{figure}


\begin{references}

\bibitem[*]{newaddress} Present address:  Department of Physics, University
of Arizona, Tucson, AZ 85721.  Electronic mail: powers@physics.arizona.edu,
gold@physics.arizona.edu.

\bibitem{CrHo} M. Cross and P. Hohenberg, Rev. Mod. Phys. {\bf 65},
(1993).

\bibitem{Plateau} J. Plateau, {\sl Statique experimentale
et theorique des liquides soumis aux seules forces moleculaires}
(Gautier-Villars, 1873).

\bibitem{Strutt} Lord Rayleigh, Proc. Lond. Math. Soc. {\bf 10}, 4 (1879);
Phil. Mag. {\bf 34}, 145 (1892).

\bibitem{BarZiv} R. Bar-Ziv and E. Moses, Phys. Rev. Lett. {\bf 73}, 1392
(1994); R. Bar-Ziv and E. Moses, unpublished.

\bibitem{Mather} P.T. Mather, K.P. Chaffee, T.S. Haddad, and J.D. Lichtenhan,
Polym. Preprints (Am. Chem. Soc.,
Div. Poly. Chem.) {\bf 37}, 765 (1996).

\bibitem{MT} E. S. Matsuo and T. Tanaka, Nature {\bf 358}, 482 (1992);
B. Barriere, K. Sekimoto, and L. Leibler,
preprint 1996.

\bibitem{VE} M. Renardy, J. Non-Newtonian Fluid Mech., {\bf 59} 267
(1995) and references therein.

\bibitem{expts} See for example W-K. Lee, K-L. Yu, and R. W. Flumerfelt,
Int. J. Multiphase Flow {\bf 7}, 385 1981.

\bibitem{slowing_down} Conservation of lipid leads to an eventual
slowing down of the front at long times \cite{GNPS}, an effect absent
here since the interfacial area is not conserved.


\bibitem{PNS} P. Nelson, T. Powers, and U. Seifert,
Phys. Rev. Lett. {\bf 74}, 3384 (1995).

\bibitem{Granek} R. Granek and Z. Olami,
J. Phys. II France {\bf 5}, 1348 (1995).

\bibitem{GNPS} R.E. Goldstein, P. Nelson, T. Powers, and U. Seifert,
J. de Phys. {\bf 6}, 767 (1996).

\bibitem{SBL} H. Stone, B. Bentley, and L. Leal, J. Fluid Mech. {\bf 173}
131 (1986).  See also the earlier work of G. I. Taylor, Proc. Roy. Soc.
A, {\bf 146}, 501 (1934).

\bibitem{AW} D.G. Aronson and H. F. Weinberger, {\it Adv. Math} {\bf 30},
33 (1978).

\bibitem{DeeL} G. Dee and J. Langer,
Phys. Rev. Lett. {\bf  50}, 383 (1983); E. Ben-Jacob, H.
Brand, G. Dee, L. Kramer, and J. Langer, Physica {\bf 14D}, 348 (1985);
W. van Saarloos, Phys. Rev. {\bf A37}, 211 (1988).

\bibitem{Saville} S. Sankaran and D.A. Saville, Phys. Fluids A {\bf 5},
1081 (1993).

\bibitem{Fish} R. Fisher, Ann. Eugenics {\bf 7}, 355 (1937);
A. Kolmogorov, I. Petrovsky, and N. Piskunov,
Bull. Univ. Moskou, Ser. Internat., Sec. A {\bf 1}, 1 (1937).

\bibitem{blowup} A. Friedman and B. McLeod, Indiana Univ. Math. J.
{\bf 34}, 425 (1985); Y. Giga and R. Kohn, Comm. Pure Appl. Math
{\bf 42}, 297 (1989), and references therein.

\bibitem{nlmsc}{We ignore for simplicity the
asymmetric potentials ${\cal V}(u)$
for which this linear
analysis does not give the correct asymptotic front speed
and shape\cite{DeeL}.}

\bibitem{Goldenfeld} F. Liu and N. Goldenfeld, Phys. Rev. A {\bf 39},
4805 (1988).

\bibitem{Tomo} S. Tomotika, Proc. Roy. Soc. Lond. {\bf A150}, 322 (1932).

\bibitem{papa} D. T. Papageorgiou, Phys. Fluids {\bf 7}, 1529 (1995).

\bibitem{second_branch} We also found a second branch of roots of the MSC
equations.
At least part of it contains unphysical MSC values, so we concentrate
on the branch shown.

\bibitem{Lamb} H. Lamb, {\it Hydrodynamics}, 6th ed.,
(Cambridge, Cambridge University Press, 1993).

\bibitem{Limat} L. Limat, P. Jenffer, B. Dagens, E. Touron, M. Fermigier,
and J.E. Wesfreid, Physica D {\bf 61}, 166 (1992).

\bibitem{Steen} M.J. Russo and P.H. Steen, Phys. Fluids A {\bf 1},
1926 (1989).

\bibitem{drain} S. J. VanHook, M. F. Schatz, W. D. McCormick, J. B. Swift,
and H. Swinney, Phys. Rev. Lett. {\bf 75}, 4397 (1995).

\end{references}
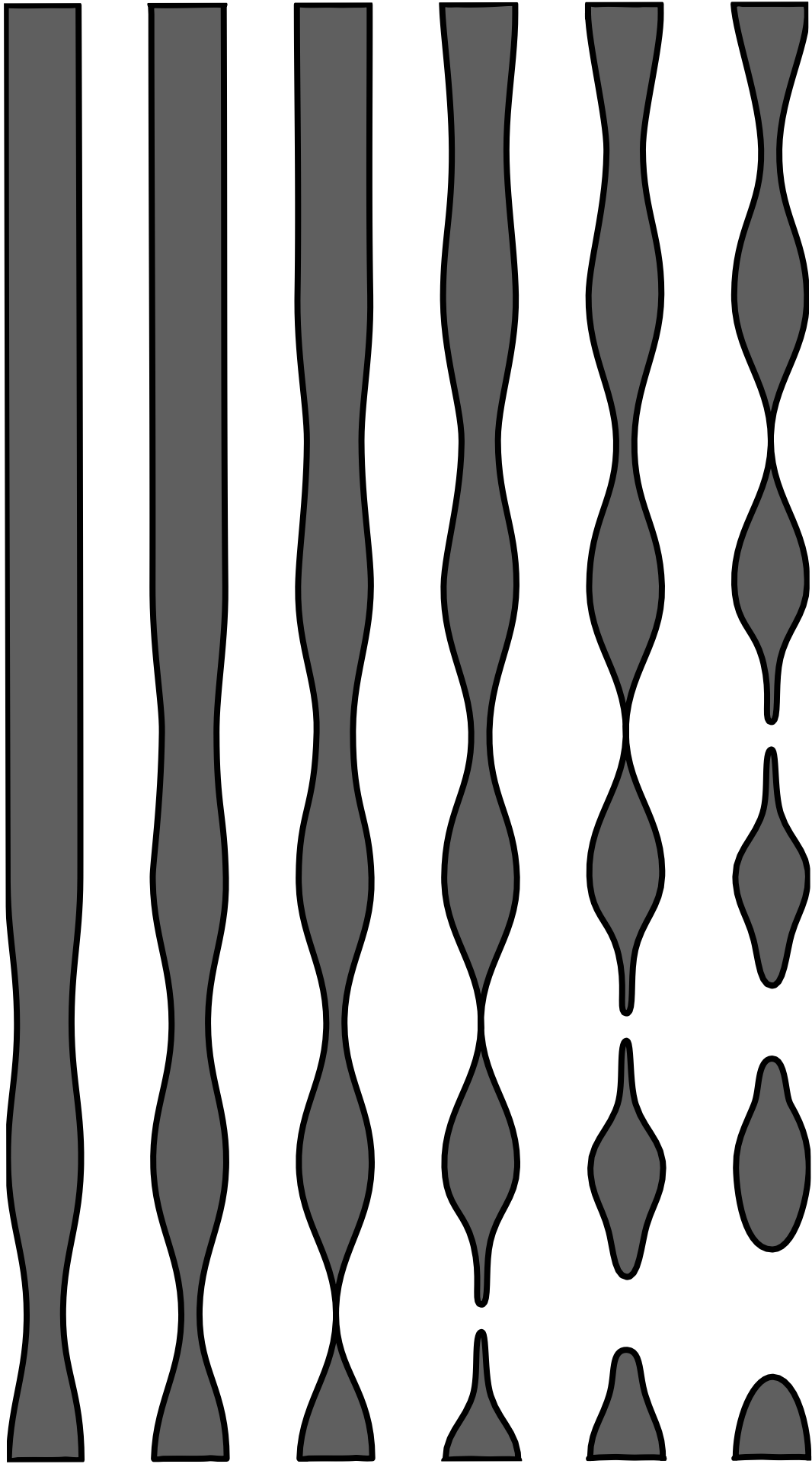
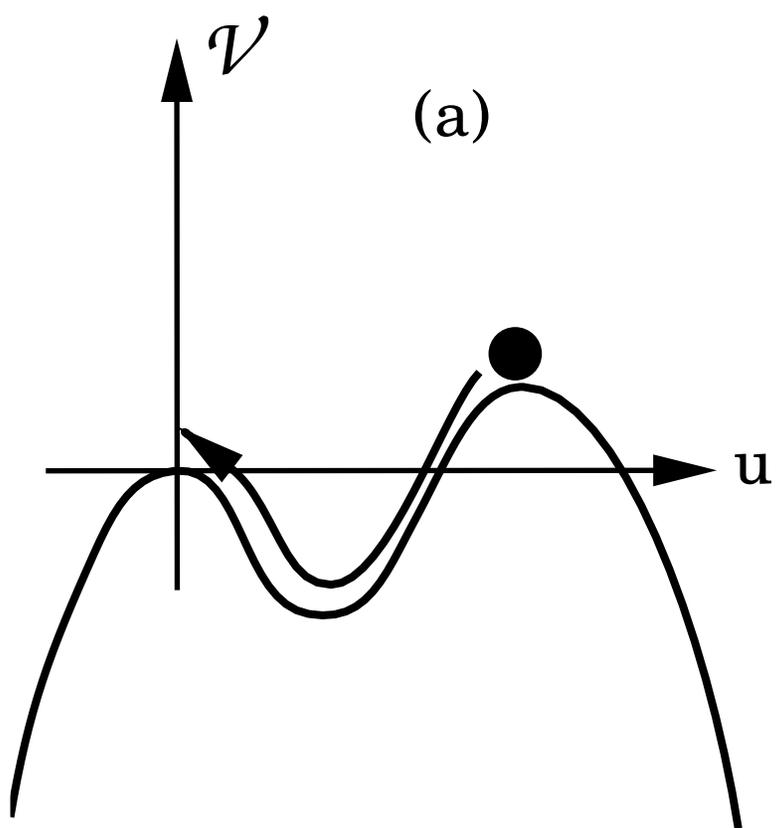
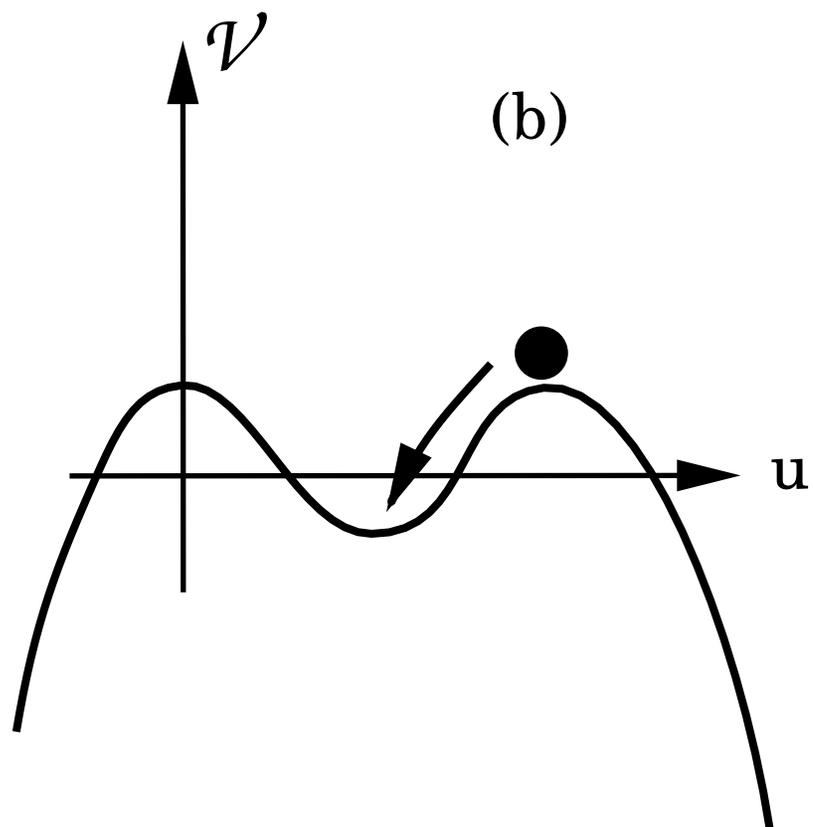
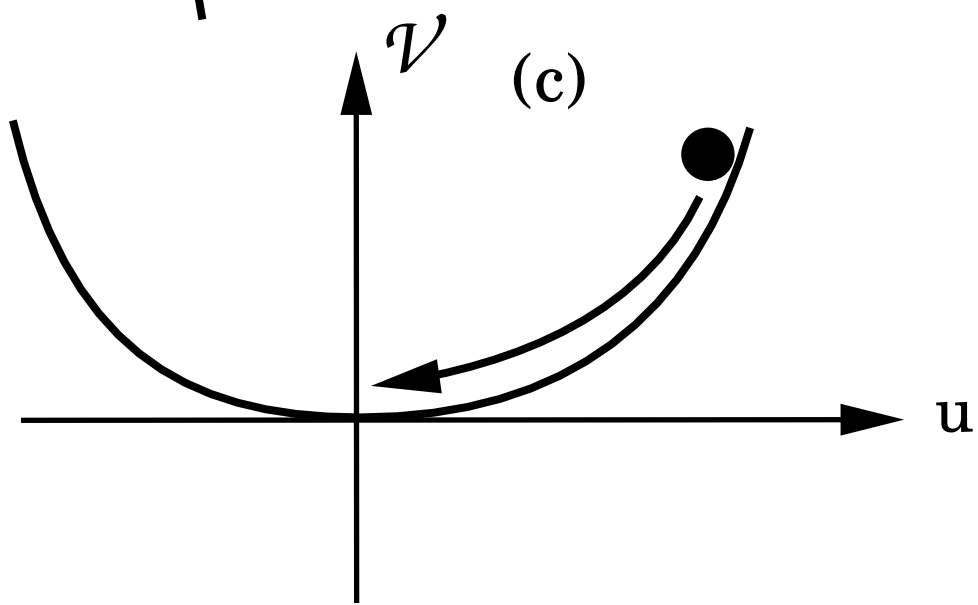
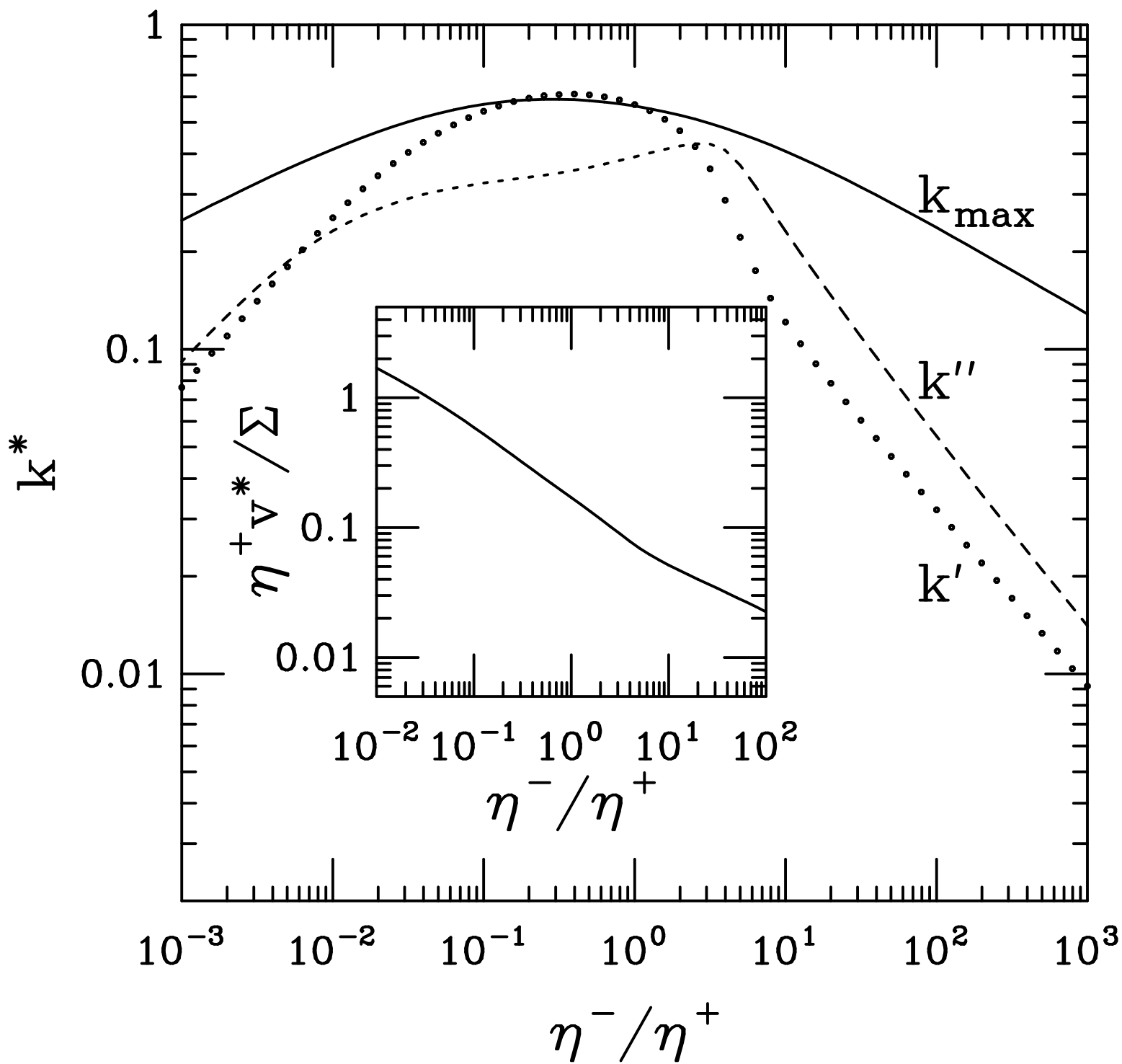
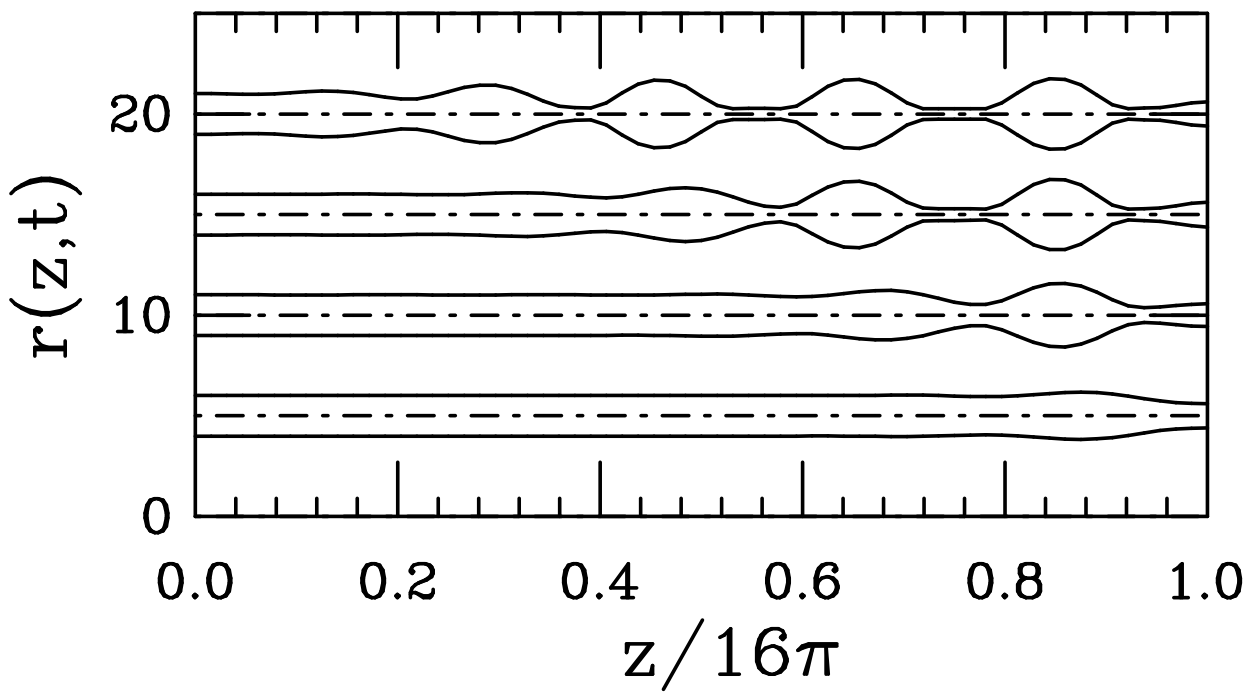
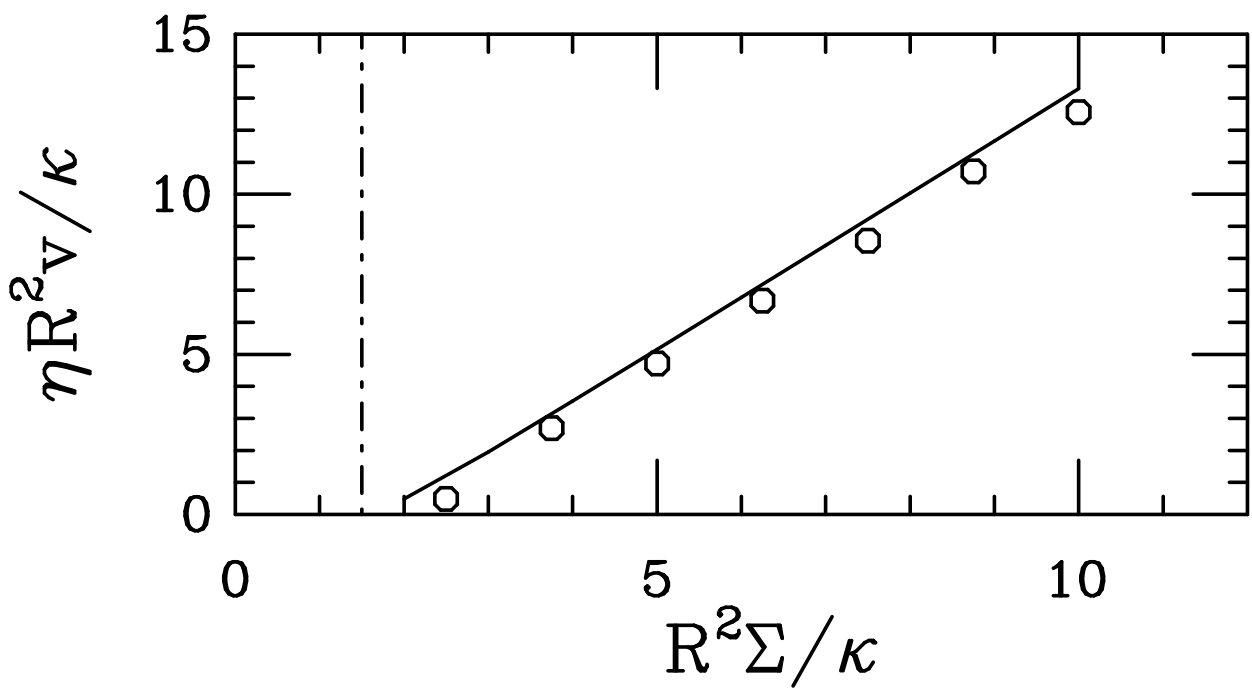
\end{document}